%% file: main.tex
\documentclass[preprintnumbers,amsmath,amssymb,pra,showkeys,11pt]{revtex4-1}
\usepackage{color}
\usepackage{graphicx}
\usepackage{hyperref}
\usepackage{cleveref}
\def\ie{i.e., }

\def\bravert{\egroup\,\vrule\,\bgroup}

{\catcode`\|=\active
  \gdef\Twoint#1{\left(\mathcode`\|"8000\let|\bravert {#1}\right)}}
{\catcode`\|=\active
  \gdef\Braket#1{\left<\mathcode`\|"8000\let|\bravert {#1}\right>}}

\newcommand{\beq}{\begin{equation}}
\newcommand{\eeq}{\end{equation}}
\newcommand{\beqa}{\begin{eqnarray}}
\newcommand{\eeqa}{\end{eqnarray}}
\newcommand{\bea}{\begin{array}}
\newcommand{\eea}{\end{array}}

\newcommand{\bef}{\begin{figure}}
\newcommand{\ef}{\end{figure}}
\newcommand{\bc}{\begin{center}}
\newcommand{\ec}{\end{center}}
\newcommand{\bt}{\begin{table}}
\newcommand{\et}{\end{table}}
\newcommand{\btb}{\begin{tabular}}
\newcommand{\etb}{\end{tabular}}

\def\etal{\mbox{et al. }}

\begin{document}
\title {\itshape{Implementation of the exact semi-classical light-matter interaction - the easy way}}

\author{Lasse Kragh S{\o}rensen$^{a}$$^{\ast}$\vspace{6pt},
        Emil Kieri$^{b}$\vspace{6pt},
        Shruti Srivastav$^{a}$\vspace{6pt},
        Marcus Lundberg$^{a}$\vspace{6pt}, and
        Roland Lindh$^{a,c}$\vspace{6pt} \\
        $^{a}${\em{Department of Chemistry - {\AA}ngstr{\"o}m Laboratory,
                   Uppsala University,
                   Box 538,
                   S-75121 Uppsala,
                   Sweden}};
         $^{b}${\em{Division of Scientific Computing, Department of Information Technology,
                    Uppsala University, 
                    POB 337,
                    SE-75105 Uppsala, 
                    Sweden}};
        $^{c}${\em{Uppsala Center of Computational Chemistry - UC$_3$,
                   Uppsala University,
                   POB 538,
                   S-75121 Uppsala,
                   Sweden}}\vspace{6pt}}

\email{lasse.kragh.soerensen@gmail.com}
\bigskip


\begin{abstract}
 \input{abstract}

\end{abstract}

\maketitle

\section{Introduction}
\label{SEC:intro}
\input{intro}

\section{Theory}
\label{SEC:theo}
\input{theory}

\section{Application}
\label{SEC:appl}
\input{app}

\section{Perspective}
\label{SEC:per}
\input{per}

\section{Conclusion}
\label{SEC:summ}
\input{summary}

\section*{Acknowledgement(s)}
\input{ack}


\bibliography{full}

\end{document}

%% file: abstract.tex








We present an analytical and numerical solution of the calculation of
the transition moments for the exact semi-classical light-matter interaction
for wavefunctions expanded in a Gaussian basis. By a simple manipulation
we show that the exact semi-classical light-matter interaction of a plane wave can 
be compared to a Fourier transformation of a Gaussian where analytical
recursive formulas are well known and hence making the 
difficulty in the implementation of the
exact semi-classical light-matter interaction comparable to
the transition dipole. Since the evaluation of the 
analytical expression involves a new Gaussian we instead have chosen to
evaluate the integrals using a standard Gau{\ss}-Hermite quadrature since this is faster.  
A brief discussion of the numerical advantages of the 
exact semi-classical light-matter interaction in comparison
to the multipole expansion along with the unphysical interpretation
of the multipole expansion is discussed.
Numerical examples on [CuCl$_4$]$^{2-}$
to show that the usual features of the multipole expansion is immediately
visible also for the exact semi-classical light-matter interaction
and that this can be used to distinguish between symmetries.
Calculation on [FeCl$_4$]$^{1-}$ is presented to demonstrate the
better numerical stability with respect to the choice of
basis set in comparison to the multipole expansion and finally 
Fe-O-Fe to show origin independence is a given for the exact operator.
The implementation is freely available in OpenMolcas.

%% file: intro.tex
Over the years a large variety of spectroscopies has been developed
which has given a great understanding of molecules and materials
from basic characterization \cite{spec}. 
All spectroscopies, until recently \cite{PhysRevLett.116.061102},
all come from the interaction between external or internal 
electromagnetic fields. 
While a great deal of information can be extracted
from experimental spectra alone, the more detailed
correspondence between observed properties and molecular
 structures is often better illuminated when experimental
results are combined with theoretical results since individual
transitions can be separated.

In reconstructing the experimental spectra from theory
it is necessary to introduce the given external electromagnetic
fields in the description of the molecular system.
The external fields used in the different types of spectroscopy are often weak in comparison
to the atomic fields, or does not significantly perturb the system before measurement,
and can therefore be treated classically \cite{spinger_handbook,Joachain2000,Posthumus} and
as a perturbation. Usually, for laser fields, 
the electromagnetic field is described by a plane wave
where the wave vector is a complex exponential function. Traditionally a multipole
expansion is introduced and
truncated at some finite order to describe the interaction of
the external electromagnetic field with the system. The first term in
this multipole expansion is the electric dipole, and the next
term that is included is typically the electric quadrupole, followed
by other magnetic and electric multipoles. While simple, the higher
order terms depend on the choice of origin for the multipole
expansion, at least in cases where there are non-zero terms of lower
order. For weak fields, which can be treated as a perturbation,
the problem of origin dependence 
was recently solved by Bernadotte \etal \cite{bernadotte2012origin}
simply by truncating the multipole in the observable wave vector
and not in the non-observable transition moments traditionally
done. A complete expansion to the second order, most commonly
associated with electric quadrupole, then requires calculations
up to magnetic quadruples and electric octupoles. 

Bernadotte \etal \cite{bernadotte2012origin} showed that origin dependence is exact when using the
velocity gauge. We later showed that origin independence in a finite
basis set can also be accomplished in the length gauge but what is typically referred to as the length gauge is actually a mixed
gauge, with the electric and magnetic components in the length and velocity gauges respectively.\cite{SORENSEN2017}  Origin
independence, in finite basis sets, is not conserved in this mixed gauge.\cite{oiqm} Furthermore the increased basis set requirement 
and convergence behaviour for every order
in the multipole expansion 
cannot be overlooked \cite{crossley,oiqm}.


An alternative way to evaluate the oscillator strengths is
to simply use the exact semi-classical light-matter interaction
and not perform any multipole expansion. In this way there
will only be one type of integrand that needs to be evaluated
and not, like for the multipole expansion, many integrands with 
different basis sets requirements. Exact semi-classical light-matter
interactions of a plane wave have been implemented
previously.\cite{Lehtola,list2015beyond,list2016average} 
The evaluation of the integrals for the exact semi-classical light-matter interaction
has been the major obstacle in the evaluation of the operator 
and often described as being very difficult
\cite{Lehtola,bernadotte2012origin}. Either a number of new recursive
relations needs to be programmed along with the need to introduce 
trigonometric functions\cite{list2015beyond,list2016average}
or a Fourier transformation of the overlap between basis function
should be performed.\cite{Lehtola} We, however, intend to show that the evaluation of the integrals for
the exact semi-classical light-matter interaction 
in a Gaussian basis set is actually very simple
and can be performed using either analytical formulas
or standard integral evaluation methods in quantum chemistry.


To illustrate the behavior of the exact operator we will perform
calculations with high-energy photons, which corresponds to large k
vectors, rapidly oscillating fields, and thus larger relative
intensity of higher-order terms in the plane-wave expansion. In X-ray
absorption spectroscopy (XAS), the K-edge of first-row transition
metals, typically associated with electric dipole-allowed $1s$ to $4p$
transitions, uses photon energies of thousands of eV. Before the
rising edge, there are weaker pre-edge transitions assigned to $1s$ to
$3d$ transitions, which provides insight into the nature of the bonding between
the transition metal(s) and ligands.\cite{shulman1976observations,hahn1982,westre1997multiplet} Since the $1s$ to $3d$
is dipole forbidden in centrosymmetric environments, higher-order terms in the 
multipole expansion must be included in order to describe
these transitions, or by using the exact operator.\cite{list2016average}

The X-ray calculations will be performed using the restricted active
space (RAS) method, which is a multiconfigurational wavefunction
approach.\cite{olsen2,malmqvist1990restricted} RAS has been successfully applied to simulate L-edge XAS and resonant
inelastic X-ray scattering (RIXS) of several transition metal
systems.\cite{Josefsson12,bokarev2013state,pinjari2014restricted}
We have also implemented the second-order expansion of the wave vector
to describe XAS and resonant inelastic X-ray scattering (RIXS) in the
K pre edge.\cite{guo_2016,guo2016molecular} 


The presented examples all represent cases with weak electromagnetic
fields. However, in the past decades with advent of very short and brilliant
laser pulses the perturbative treatment can break down in and one enters
the strong field regime where the external and atomic field must treated on 
equal footing and a dynamical treatment is necessary \cite{spinger_handbook,Joachain2000,Posthumus}. For strong fields,
beyond the dipole approximation, the problem of origin dependence
still persists for the multipole expansion. We will here allude
to how the work on the exact semi-classical light-matter interaction
can be carried directly over to the strong field regime because all
interaction terms can be evaluated using the same simple integrals. This also means that the
method also could be used in simulations of dynamics 
of molecules in strong electric and magnetic fields and this, we believe, is where
the real strength of the approach may lie.

For self consistency we will in Sec. \ref{perturb} recapitulate
the perturbative treatment of molecules in weak electromagnetic fields
and the multipole expansion. Thereafter in Sec. \ref{eva} we will show how
the integrals for the exact semi-classical light-matter interaction
can be evaluated using standard quantum chemistry integral programs.
Followed by
the isotropically averaged oscillator strengths in Sec. \ref{iso_ave}
For the applications in Sec. \ref{SEC:appl} we demonstrate the advantage
of using the exact semi-classical light-matter interaction
instead of the multipole expansion on different systems
which has been problematic with the multipole-expansion approach.
A perspective on
and the possibility of dynamics simulations with
the exact semi-classical light-matter interaction is given
in Sec. \ref{real_time} and finally a summary and conclusion
in Sec \ref{SEC:summ}. 


%% file: theory.tex
In the first of the two parts of this section we will briefly discuss
the well-known formulas for the semi-classical light-matter interaction
and how the oscillator strengths usually are calculated
from perturbation theory along with a short discussion
of the unphysical interpretation of the multipole expansion often seen.
In the second part we will show the integrals for
the exact semi-classical light matter interaction
can be evaluated analytically along with an
easy way to compute the integrals
using a standard Gau{\ss}-Hermite quadrature. 
Finally the isotropic averaging of the
exact semi-classical light matter interaction
is mentioned.

\subsection{Perturbation from weak fields}
\label{perturb}

Throughout this section it is assumed that the electromagnetic fields are weak
and therefore can be treated as a perturbation of the molecular system.
The zeroth order Hamiltonian, in our case, is the Schr\"odinger equation 
within the Born-Oppenheimer approximation
\beq
\label{tindep}
\hat H_0 = \sum_{i=1}^{N} \frac{\bm{\hat p}_i^2}{2 m_e} + V(\bm{r}_1, \ldots , \bm{r}_N)
\eeq
which is exposed to a time-dependent perturbation $\hat U(t)$
\beqa
\label{tinde}
\hat U(t) &=& - \frac{e}{m_e c} \sum_i \bm{A(r}_i,t) \bm{\cdot \hat p}_i 
              + \frac{e^2}{2 m_e c^2} \bm{A^2(r}_i,t)
              - \frac{ge}{2 m_e c} \sum_i \bm{B(r}_i,t) \bm{\cdot \hat s}_i \\
\label{tinde1}
          &=& \frac{e A_0}{2 m_e c} \sum_i \Big[ 
exp(\imath (\bm{k \cdot r}_i -\omega t))( \bm{\mathcal{E} \cdot \hat p}_i)   \\
\label{tinde2}
          &+& \frac{e A_0}{4 c} (exp(2 \imath (\bm{k \cdot r}_i -\omega t)) + 1) \\
\label{tinde3}
&+& \imath \frac{g}{2} exp(\imath \bm{k \cdot r}_i -\omega t)( \bm{k \times \mathcal{E}}) \cdot \bm{\hat s}_i + c.c. \Big]
\eeqa
from a monochromatic linearly polarized electromagnetic wave where
$\bm{k}$ is the wave vector pointing in the direction of 
propagation, $\bm{\mathcal{E}}$ the polarization vector
perpendicular to $\bm{k}$, $\omega$ is
the angular frequency, $\bm{\hat s}$ the spin and $A_0$ the amplitude of the vector potential. 

Of the terms in Eqs. \ref{tinde1}-\ref{tinde3} often 
the dipole approximation is taken
meaning that only the zeroth order term
in the vector potential $(\bm{A})$ in Eq. \ref{tinde1} is included
\beq
\label{multipole}
exp(\imath \bm{k \cdot r}_i ) = 1 + \imath (\bm{k \cdot r}_i ) - \frac{1}{2}(\bm{k \cdot r}_i )^2 + \ldots
\eeq
While the dipole approximation suffices for optical
transitions for analyzing the K-edge in X-ray spectroscopy
terms up to second order must be included.
The $(\bm{A}^2)$ in Eq. \ref{tinde2} is mostly relevant for
strong fields and will always depend explicitly on the field
strength $A_0$ which makes little sense for weak fields, treated perturbatively, where this
dependence is removed from the terms in Eqs. \ref{tinde1} and \ref{tinde3}.
Eq. \ref{tinde3} describes the interaction between the
spin and the magnetic field and is relevant when describing
open shell transitions.
Furthermore the values of all terms in Eqs. \ref{tinde1}-\ref{tinde3} also
depend on the choice of gauge, though the sum is constant. 
In the Coulomb gauge, which is the usual choice in molecular physics,
$(\bm{A}^2)$ has a minimum \cite{gubarev2001} and will be neglected
in the applications.

Using Fermi's golden rule transitions
only occur when the energy difference between the eigenstates of the unperturbed molecule
matches the frequency of the perturbation
\beq
\omega = \omega_{0n} = \frac{E_n - E_0}{\hbar}
\eeq
and the explicit time dependence can be eliminated from the transition rate
\beq
\label{trarate}
\Gamma_{0n}(\omega) = \frac{2 \pi }{\hbar} 
| \langle 0 | \hat U | n \rangle |^2 \delta(\omega-\omega_{0n}) =
\frac{ \pi A_0^2}{2 \hbar c} 
| T_{0n} |^2 \delta(\omega-\omega_{0n})
\eeq
and the effect of the weak electromagnetic field can now
be expressed as a time-independent expectation value.
From Eq. \ref{trarate} the relation between the transition moments $T_{0n}$  
and the time-independent part of $\hat U$ in Eq. \ref{tinde} is seen. 
From the transition moments $T_{0n}$ the oscillator strengths $f_{0n}$
\beq
\label{oscillator}
f_{0n} = \frac{2 m_e}{e^2 E_{0n}} | T_{0n}|^2 ,
\eeq
where $E_{0n} = E_n - E_0$ is the difference in the energy of the
eigenstates of the unperturbed molecule,
can then be calculated. The amplitude of the
electric and magnetic field $E_0 = B_0 = A_0 k$ or intensity
therefore does not have to be defined for Eqs. \ref{tinde1} and \ref{tinde3}
while for the quadratic $\boldmath{A}$ term in 
Eq. \ref{tinde2} the amplitude is still needed. 

Traditionally a multipole expansion of the exponential
function of the perturbation in Eq. \ref{tinde} is performed
which gives rise
to the non-observable electric and magnetic dipole, quadrupole and higher order approximations
for the transition moments $T_{0n}$. Unfortunately
such an expansion in the transition moments $T_{0n}$ is
only origin independent for the dipole and in the
limit of a complete expansion.

Origin independence, however, 
appears naturally provided that the collection of the terms in 
Taylor expansion of the exponential of the 
wave vector {\boldmath{$k$}} in Eq. \ref{tinde}
are collected to the same order in the observable oscillator strengths
\beq
\label{osstr}
f_{0n} = f_{0n}^{(0)} + f_{0n}^{(1)} + f_{0n}^{(2)} + \ldots = \frac{2 m_e}{e^2 E_{0n}} |T_{0n}^{(0)} + T_{0n}^{(1)} + T_{0n}^{(2)} + \ldots |^2
\eeq
as shown by Bernadotte \etal \cite{bernadotte2012origin}.
Lestrange \etal \cite{Lestrange} demonstrated that collecting the terms in the
oscillator strengths according to Eq. \ref{osstr}
does not always ensure a positive total oscillator strength
when truncating the expansion since
the perfect square of the transition moments is
broken. The total
negative oscillator strengths when truncating Eq. \ref{osstr}
appear to be a basis set problem that can occur for
unbalanced basis sets for some transitions \cite{oiqm}.

While the truncation in the oscillator strength 
eliminates the problem of origin dependence the
multipole expansion, however, introduces an 
increasing demand on the basis set for every
order in the expansion since the integrand
changes for every order \cite{crossley}. This means 
that in order to calculate the K pre-edge peaks in an
X-ray spectrum the basis set must be able to 
accurately describe all terms at least up to second
order in the transition moment \ie the electric
octupole and magnetic quadrupole terms. While 
the higher order terms could be expected to be small
these can be grossly overestimated 
in some basis sets \cite{oiqm}.

Lastly, none of the terms in the multipole expansion in Eq. \ref{osstr}
are individually observable so there is no physical
argument to perform and try to interpret the expansion
except that this is what is historically done.
The same argument also goes for the 
origin independent oscillator strength \cite{bernadotte2012origin}
which despite the origin independence 
still are not individually observable. 
Therefore trying to interpret spectra in terms
of the different orders in the multipole expansion or even
as electric or magnetic is completely unphysical since
only the total can be observed and changes of coordinate
system can significantly alter the interpretation \cite{oiqm}.

\subsection{Evaluation of the integrals for the exact semi-classical light-matter interaction}
\label{eva}

The evaluation of the integrals for the exact semi-classical light-matter interaction
has been the major obstacle in the evaluation of the operator.\cite{Lehtola,bernadotte2012origin}  We will show that the
exact semi-classical light-matter interaction of a plane wave can
be thought of as a Fourier transformation of the
overlap between basis functions and that this can 
be solved analytically. In the Gaussian basis sets
we use this just results in a new Fourier-transformed Gaussian.
The evaluation of the integrals
are therefore very similar to those found
for the overlap and operators in a Gaussian Planewave basis set \cite{polasek,polasek2,molnar}
and similarities are shared with the plane wave representations
of the electromagnetic field \cite{devaney1974}. 

In order to evaluate Eq. \ref{trarate} for the perturbation in
Eq. \ref{tinde} the matrix element
\beq
\label{traper}
\langle 0 | \hat U | n \rangle = \sum_{\mu \nu} U_{\mu \nu}^{AB} \gamma_{\mu \nu}^{AB}
\eeq 
must be calculated. In Eq. \ref{traper} $U_{\mu \nu}^{AB}$ is the 
integral matrix for the orbital bases $A$ and $B$ with indices $\mu$ and $\nu$
and likewise defined for the transition density matrix $\gamma_{\mu \nu}^{AB}$ \cite{Malmqvist}.
For a wave function expanded in 
Gaussians the individual terms in  $U_{\mu \nu}^{AB}$ from 
Eq. \ref{tinde1} correspond to evaluating
integrals of the form
\beq
\label{theint}
I = \langle \chi_{\mu} | e^{\pm \imath \bm{k \cdot r}} \bm{\hat p} | \chi_{\nu}  \rangle \bm{ \cdot \mathcal{E}}
\eeq
where the real-valued atomic Cartesian basis functions $\chi_{\mu}$ and $\chi_{\nu}$
are expressed as
\beqa
\chi_{\mu}(\bm{r}) &=& \chi_{i,j,k}(\bm{r}, \alpha_{\mu}, \bm{A}) \\
                   &=& (x-A_x)^{i} (y-A_y)^j (z-A_z)^k e^{-\alpha_{\mu} \Vert \bm{r} - \bm{A} \Vert ^2} \\
                   &=& \chi_i (x,\alpha_{\mu},A_x) \chi_j (y,\alpha_{\mu},A_y) \chi_i (z,\alpha_{\mu},A_z)
\eeqa
in their different components, where $i$, $j$, and $k$ represent the order of the Cartesian components $x$, $y$, and $z$, respectively. The integral in Eq. \ref{theint}
can be factorized into three one-dimensional integrals
\beq
I_x = \int_{-\infty}^{\infty} \chi_i (x,\alpha_{\mu},A_x) e^{\pm \imath k_x x} \epsilon_x \hat p_x \chi_j (x,\alpha_{\nu},B_x) dx.
\eeq
Applying the differentiation operator $\hat p_x = - \imath \hbar \frac{\partial}{\partial x}$
we find
\beq
\label{twoterm}
I_x = - \imath \hbar \epsilon_x \int_{-\infty}^{\infty} 
\chi_i (x,\alpha_{\mu},A_x) e^{\pm \imath k_x x} \left(j \chi_{j-1} (x,\alpha_{\nu},B_x) - 2 \alpha_{\nu} \chi_{j+1} (x,\alpha_{\nu},B_x)\right) dx
\eeq
that the integral $I_x$ can be expressed as a sum of two
terms. From Eq. \ref{twoterm} it is seen that both
terms are of the form
\beq
\label{form}
\int_{-\infty}^{\infty} e^{\pm \imath k_x x} \chi_i (x,\alpha_{\mu},A_x) \chi_j (x,\alpha_{\nu},B_x).
\eeq
Using the Gaussian product formula we see that the expression in Eq. \ref{form} 
is akin to a Fourier transformation of a Gaussian
from real space $x$ to $k_x$ space. Integrals of the from in Eq. \ref{form} can be solved analytically using
recursive formulas for the analytical Fourier representation of Gaussians \cite{polasek}.
Eq. \ref{form} can also be viewed as the Fourier transformation of
the overlap between two basis functions as also noted by Lethola \etal \cite{Lehtola}.

Since the Fourier transformation of a Gaussian is a new Gaussian 
we have chosen not to use the analytical form but instead 
rewrite the integral in Eq. \ref{form} to a form which easily can be evaluated
by a standard Gau{\ss}-Hermite quadrature. 
Using the Gaussian product formula on Eq. \ref{form}
\beqa
I_x'  &=& \int_{-\infty}^{\infty} e^{\pm \imath k_x x} \chi_i (x,\alpha_{\mu},A_x) \chi_j (x,\alpha_{\nu},B_x) \\
\label{simgp}
      &=& e^{-\frac{\alpha_{\mu} \alpha_{\nu}}{\zeta} (A_x - B_x)^2} \int_{-\infty}^{\infty} 
          (x-A_x)^{i} (x-B_x)^{j} e^{-\zeta (x-P_x)^2 \pm \imath k_x x} dx
\eeqa
where $\zeta = \alpha_{\mu} + \alpha_{\nu}$ and $P_x = (\alpha_{\mu} A_x + \alpha_{\nu} B_x)/ \zeta$
we can complete the square in the exponent
\beq
\label{square}
I_x' = e^{-\frac{\alpha_{\mu} \alpha_{\nu}}{\zeta} (A_x - B_x)^2} e^{\gamma}  \int_{-\infty}^{\infty}
       (x-A_x)^{i} (x-B_x)^{j} e^{-\zeta (x-Q_x)^2} dx
\eeq
where $Q_x = P_x \pm \imath k_x/(2 \zeta) $ and $\gamma = \zeta (Q_x^2 - P_x^2)$.
We here notice that for mixed Gaussian Plane wave basis set
expressions similar to Eq. \ref{simgp} for an overlap appears \cite{molnar}.

Making a change of variables $z=\sqrt{\zeta} (x - Q_x)$ the integral in Eq. \ref{square}
can now be transformed to 
\beq
\label{expand}
I_x' = \Theta \lim_{R\rightarrow \infty} \int_{-z=\sqrt{\zeta} (R - Q_x)}^{z=\sqrt{\zeta} (R - Q_x)} 
       (\frac{z}{\sqrt{\zeta}} + Q_x - A_x)^{i} (\frac{z}{\sqrt{\zeta}} + Q_x - B_x)^{j} e^{-z^2} dz
\eeq
where 
\beq
\Theta = e^{-\frac{\alpha_{\mu} \alpha_{\nu}}{\zeta} (A_x - B_x)^2} e^{\gamma} / \sqrt{\zeta}.
\eeq
Defining the polynomial
\beq
f(z) = \Theta (\frac{z}{\sqrt{\zeta}} + Q_x - A_x)^{i} (\frac{z}{\sqrt{\zeta}} + Q_x - B_x)^{j}
\eeq
Eq. \ref{expand} can be written a little more compact
\beq
\label{compact}
I_x' = \lim_{R\rightarrow \infty} \int_{-z=\sqrt{\zeta} (R - Q_x)}^{z=\sqrt{\zeta} (R - Q_x)}
       f(z) e^{-z^2} dz.
\eeq
Since the integral in Eq. \ref{compact} is analytic the integration is independent
of the path and can therefore be split into
\beqa
I_x' &=& \lim_{R\rightarrow \infty} \int_{-z=\sqrt{\zeta} (R - Q_x)}^{z=\sqrt{\zeta} R} f(z) e^{-z^2} dz \\
     &+& \lim_{R\rightarrow \infty} \int_{-z=\sqrt{\zeta} R}^{z=\sqrt{\zeta} R} f(z) e^{-z^2} dz \\
     &+& \lim_{R\rightarrow \infty} \int_{-z=\sqrt{\zeta} R}^{z=\sqrt{\zeta} (R - Q_x)} f(z) e^{-z^2} dz.
\eeqa
Since $\sqrt{\zeta} > 0 $ and the exponential decay of the integrand as
$\mathfrak{R}_z \rightarrow \pm \infty$ two of the integrals vanishes
leaving
\beq
I_x' = \lim_{R\rightarrow \infty} \int_{-z=\sqrt{\zeta} R}^{z=\sqrt{\zeta} R} f(z) e^{-z^2} dz 
     = \int_{-\infty}^{\infty} f(z) e^{-z^2} dz
\eeq
for which the Gau{\ss}-Hermite quadrature is designed
to compute. Since $z$ is complex the Gau{\ss}-Hermite quadrature
must use complex numbers. With the standard Gau{\ss}-Hermite 
nodes $z_n$ and weights $w_n$, we compute the integral as
\beq
I_x' = \sum_n w_n f(z_n)
\eeq
or equivalently with the transformed quadrature nodes $x_n = z_n / \sqrt{\zeta} + Q_x$
\beq
I_x' = \Theta \sum_n w_n (x_n - A_x)^{i} (x_n - B_x)^j.
\eeq
The total integral in Eq. \ref{theint} can therefore simply be written as
\beq
I = I_x' * I_y' * I_z'.
\eeq

Due to the
similarities between the electric term in the exact semi-classical
light-matter interaction for a plane wave (Eq. \ref{tinde1}), with the
quadratic and magnetic terms (Eqs. \ref{tinde2} and \ref{tinde3}) all
these integrals can be evaluated in exactly
the same manner. All three terms are therefore programmed in OpenMolcas \cite{OpenMolcas}.
The coupling between the magnetic field and the spin in Eq. \ref{tinde3} is only non-zero
when the spin-orbit operator in the RASSI module \cite{rassi_so} is used. Eq. \ref{tinde2}
also gives a constant non-zero contribution in all directions but since Eq. \ref{tinde2} still
depends explicitly on the field strength we have neglected this term since
for the field strengths needed for Eq. \ref{tinde2} to be influential the
perturbative treatment of the light-matter interaction will break down.

The resulting formulas are not surprisingly
like those found using Rys quadrature in a Gaussian plane wave basis set 
as derived by {\v C}arsky and Pol{\'a}{\v s}ek \cite{polasek2}
and does not require any new recursive relations or
expansion in trigonometric functions like previous implementations \cite{list2015beyond,list2016average}.

\subsubsection{Isotropically averaged oscillator strengths}
\label{iso_ave}

For the terms in the multipole expansion well known
isotropically tensor averaged oscillator strengths
can be found in literature \cite{barron}.
For the exact expression no closed formula exists.
Lebedev and co-workers \cite{lebedev1,lebedev2,lebedev3,lebedev4,lebedev5,lebedev6} have devised a way of
distributing quadrature points over a unit sphere 
defining a
Lebedev grid which gives the propagation directions included
in the numerical integration for the incoming light.
By averaging over two orthogonal polarization directions 
for the different directions for the propagation the
exact isotropic average can be systematically
approximated. List \etal \cite{list2016average} have
shown that this converges very rapidly with the
number of quadrature points and we therefore
have also adopted the Lebedev grid for the isotropic averaging.

%% file: app.tex
In this section we will study the metal K pre edge XAS of two molecular systems,
[CuCl$_4$]$^{2-}$ and
[FeCl$_4$]$^{1-}$, as well as the iron dimer model complex [Fe$_2$O]$^{4+}$, to highlight
properties specific to the use of the 
exact semi-classical operator v.s. standard multipole techniques. In a classical experiment, the angular dependence of the pre-edge
intensity of single-crystal [CuCl$_4$]$^{2-}$ was used to identify the
electric quadrupole contribution and to identify the symmetry of the
singly-occupied 3d orbital. In general, the assignment of transitions
to different multipole contributions has helped to connect spectra to
the electronic structure. With the
[CuCl$_4$]$^{2-}$ example, we will demonstrate how the exact
operator reproduces the behavior of what is traditionally
referred to as an electric quadrupole transition.


As mentioned above, electric quadrupole transitions are
origin-dependent if the electric dipole contributions are
non-zero. [FeCl$_4$]$^{1-}$ has tetrahedral symmetry, and the
non-centrosymmetric ligand environment leads to intense dipole
contributions and strong contributions from many terms in the full
second-order expansion.\cite{bernadotte2012origin,guo_2016,oiqm} For some
basis sets, the expansion even leads to unphysical negative oscillator
strengths.\cite{Lestrange,oiqm,SORENSEN2017} These examples are
revisited with the exact semi-classical operator to show its stability
in incomplete basis sets.


Finally, we address an iron dimer where there is no natural choice of
the origin for the multipole expansion of an iron-centered transition. We
 show the origin independence of the exact operator by comparing the results for [Fe$_2$O]$^{4+}$ to previous
calculations using the multipole expansion in the mixed gauge.
However, before that we describe the computational
details.

\subsection{Computational details}

The geometry of the [CuCl$_4$]$^{2-}$ is taken from the X-ray crystal
structure.\cite{hahn1982} The complex has a square planar geometry, formal D$_2$h symmetry, with Cu-Cl bond lengths of
2.233 and 2.268 {\AA} and Cl-Cu-Cl angles of 89.91$^{\circ}$ and
90.09$^{\circ}$. The short bonds were placed along the x-axis. To show the effect
of the angles, another calculation in D$_{2h}$ symmetry with 
Cl-Cu-Cl angles of 90$^{\circ}$ and with all bonds along
the x- and y-axis, here labelled D$_{2h\perp}$,  were also
performed. Finally, calculations were made in D$_{4h}$ symmetry using an
average bond length of 2.2505 {\AA}. 

The geometry of [FeCl$_4$]$^{1-}$ is also taken
from an X-ray structure.\cite{warnke2010structural} The ligand
environment is tetrahedral ($T_d$ point group) with four Fe-Cl distances of 2.186
{\AA}. The geometry of Fe-O-Fe is taken from a BP86/6-311(d) geometry
optimization of  [(hedta)FeOFe(hedta)], which gives $C_{2v}$ symmetry, Fe-O distances of 1.76
{\AA} and an angle of 148 degrees.\cite{oiqm}

Orbital optimization is performed using state-average RASSCF, with
separate optimizations for ground and core-excited states as implemented in OpenMolcas.\cite{OpenMolcas} In all
calculations the metal 1s orbitals are included in RAS1, constraining
to at most one hole. For the calculations of the core-excited states,
the weights of all configurations with fully occupied 1s orbitals have
been set to zero. To avoid orbital rotation, i.e., the hole appears in
a higher-lying orbital, the 1s orbitals have been frozen in the
calculation of the final states.

[CuCl$_4$]$^{2-}$ is a formal 3d$^{9}$ complex with a singly
occupied 3d$_{x2-y2}$ orbital, leading to a doublet ground state. We
focus only on the 1s$\rightarrow$3d$_{x2-y2}$ transition and use a small RAS2 space
including seven electrons in four metal-centered orbitals, see Figure
\ref{active_space_fig}. Due to weak spin-orbit coupling, only final states of the same spin
multiplicity as the ground state are considered in the
calculations. For [CuCl$_4$]$^{2-}$ only one doublet core excited
state is necessary to include.

\bef[h!]
\bc
\includegraphics[width=0.5\textwidth]{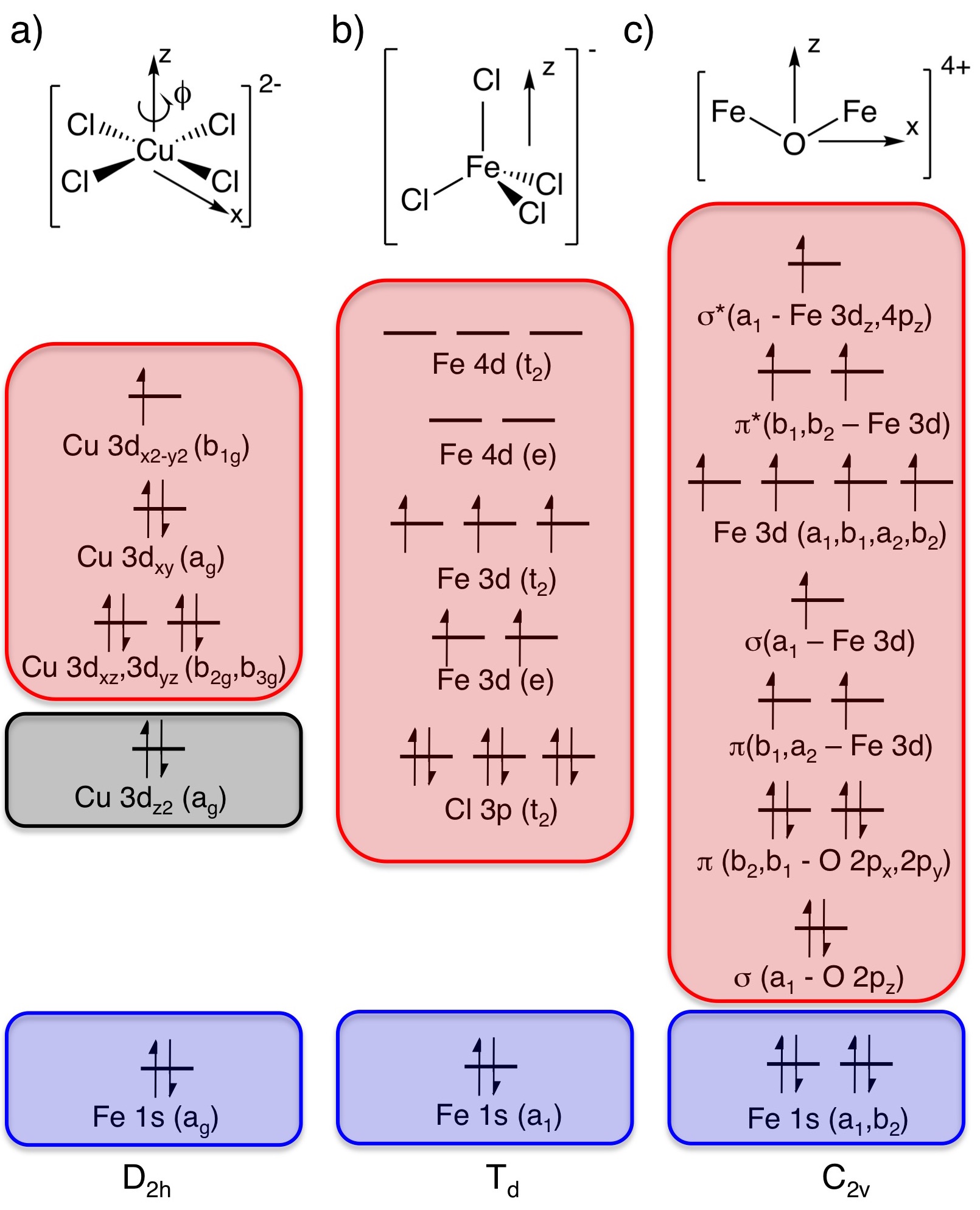}
\caption{Ligand geometries and active spaces for metal K pre-edge
  XAS modeling of a) [CuCl$_4$]$^{2-}$, b) [FeCl$_4$]$^{1-}$, and c) [Fe$_2$O]$^{4+}$.}
\label{active_space_fig}
\ec
\ef 

[FeCl$_4$]$^{1-}$ is a high-spin 3d$^{5}$ complex with a sextet ground
state. The calculations are similar to
those laid out in previous works,\cite{oiqm,SORENSEN2017}
with eleven electrons in 13 orbitals in RAS2, see Figure
\ref{active_space_fig}. The orbitals of the 
sextet excited states were averaged over 70 states. 

The ground state of the iron dimer [Fe$_2$O]$^{4+}$ is an
singlet, with five unpaired electrons on
each ferric iron coupled antiferromagnetically. To facilitate RASSCF convergence, calculations are
instead performed with ferromagnetic coupling, giving
undectet states. The RAS2 space consists of the three 2p orbitals 
of the bridging oxygen and the ten 3d orbitals of the irons, which
gives a total of 16 electrons in 13 orbitals, see Figure
\ref{active_space_fig}. 60 core-excited states were used, exactly like in previous work.\cite{oiqm} 

For the correlation treatment all calculations will be at the RASSCF
level as inclusion of dynamical correlation on the behavior of the
transitions can be assumed to be minor. Scalar relativistic effects have been included by using a second-order
Douglas-Kroll-Hess
Hamiltonian in combination with the ANO-RCC-VTZP basis
set.\cite{douglas,Hess:1986,ano-rcc_fe,ano-rcc_cl}
This basis set have been shown to perform reasonably well for both
electronic structure and for the transition moments.\cite{oiqm} The intensities for the exact operator and 
the quadrupole intensities in the mixed gauge are implemented in 
the RASSI program \cite{malmqvist1989casscf,rassi_so}
and distributed freely in the OpenMolcas package\cite{OpenMolcas}.  Simulated spectra are plotted using a
Lorentzian lifetime broadening with a full-width-at-half-maximum
(FWHM) of 1.25 eV and further convoluted with a Gaussian experimental
broadening of 1.06 eV.



\subsection{Assignment of K pre edge XAS contributions for [CuCl$_4$]$^{2-}$}

Metal K pre-edges are weak transitions on the low-energy side of the
rising edge. They are typically assigned to 1s$\rightarrow$3d
transitions. In centrosymmetric geometries these are electric dipole
forbidden and only gain intensity through what is typically referred
to as electric quadrupole transitions. However, vibronic coupling with
normal modes that break centrosymmetry allows for electric dipole
contributions also for complexes with formal centrosymmetry.

In single crystals the orientation of the molecule with respect to the
beam can be controlled. The angular dependence of the normalized peak heights in the Cu K pre edge of
[CuCl$_4$]$^{2-}$, taken from reference \cite{hahn1982}, is shown in
Figure \ref{cucl_fig}a. The angle $\phi$ shows rotation around the
molecular z-axis, with 0$^{\circ}$ representing the direction of the
electromagnetic k vector relative the short Cu-Cl bond. The
electric quadrupolar contribution is distinguished by a four-fold
periodicity of the cross section. The highest intensity is observed
for orientations bisecting the Cu-Cl bond, which makes it possible to
assign the half-filled orbital to be 3d$_{x2-y2}$.\cite{hahn1982} The isotropic
contribution is assigned to an electric dipole contribution
that gains intensity through vibronic coupling. 

\bef[h!]
\bc
\includegraphics[width=1.0\textwidth]{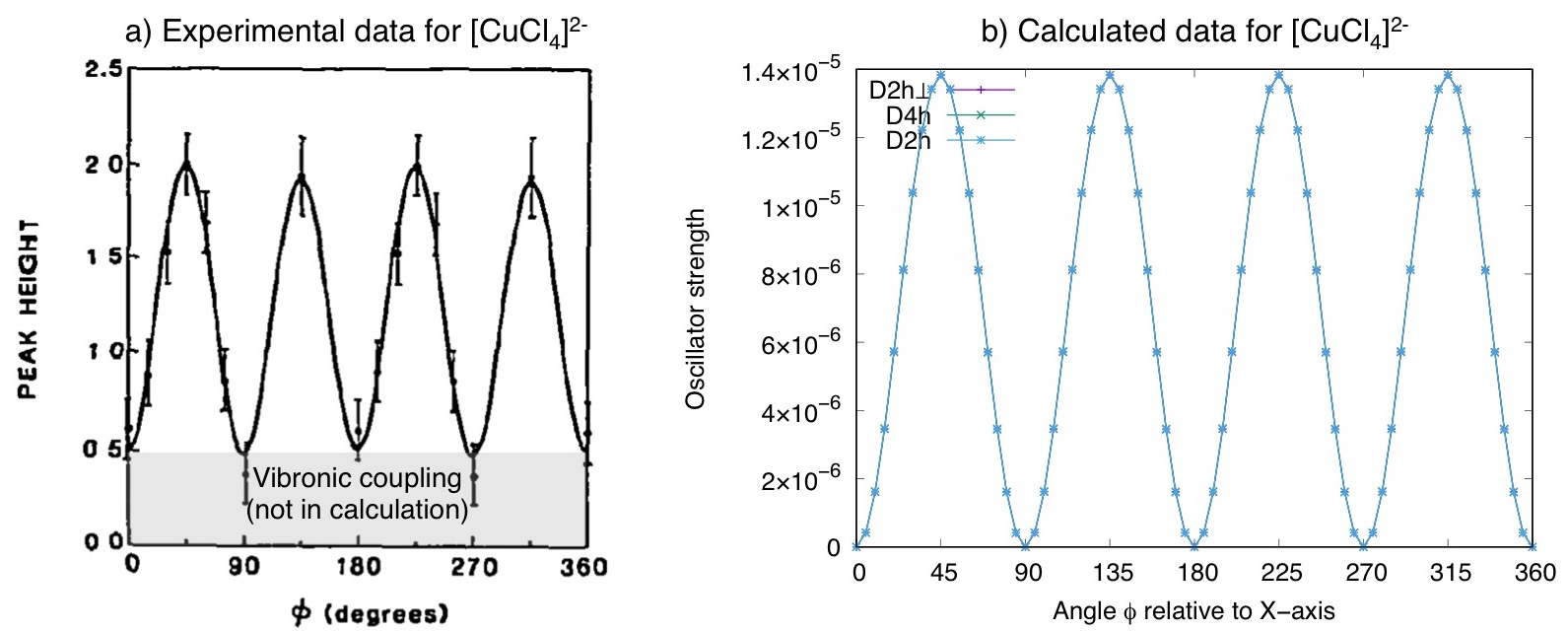}
\caption{Experimental and calculated angular dependence of the K
  pre-edge transition in [CuCl$_4$]$^{2-}$. a) Normalized K pre-edge
  peak heights from reference \cite{hahn1982}. Reproduced with
  permission from b) Oscillator strengths
  calculated using the exact semi-classical light-matter interaction
  in different in with different symmetries.}
\label{cucl_fig}
\ec
\ef 

The angular dependence of the oscillator strengths calculated using
the exact operator is shown in Figure \ref{cucl_fig}b. The four-fold
periodicity is reproduced, which is a simple illustration that the
exact operator reproduces the observables that have been traditionally
used to assign transitions to different multipole components. The
isotropic contributions are missing from the calculated spectra simply
because  vibronic coupling is not taken into account
in our calculations.



In the real complex the Cu-Cl bonds are not perfectly
symmetric. In order to see the very slight asymmetry in the angular spectrum
of [CuCl$_4$]$^{2-}$ one needs to explicitly compare the 
oscillators strengths on both sides of the peaks, e.g., 
130$^{\circ}$ and 140$^{\circ}$. Due to the  lack of 
data points and reasonably large error bars in the
experiment distortions from D$_{4h}$ are difficult to quantify in the
experimental spectrum. In Table \ref{cucl_tab}
the peak and minimum along with their neighboring values are listed
to show the asymmetry in the spectrum of [CuCl$_4$]$^{2-}$ is and how little the
values changes with nuclear geometry. From Table \ref{cucl_tab}
it is seen that the difference between the points next to 
the peak and minimum is a mere $1.0*10^{-8}$ for D$_{2h}$ which of course
is much lower than the accuracy of the calculation but still
above numerical noise. For D$_{4h}$ and the difference
between symmetrically placed points is negligible and
a numerical zero is observed at 90$^{\circ}$ as it should be.
Comparing the values for D$_{2h}$, D$_{4h}$ and D$_{2h\perp}$ directly
the difference is still below the accuracy of the calculation
and discerning between D$_{4h}$and D$_{2h\perp}$ is not possible with
the geometry differences here chosen.



\bt
\btb{cccc}
Angle & D$_{2h}$ & D$_{4h}$ & D$_{2h\perp}$ \\ \hline
130 & 0.13423083E-04 & 0.13417817E-04 & 0.13415988E-04 \\
135 & 0.13832984E-04 & 0.13834993E-04 & 0.13833106E-04 \\
1400 & 0.13408657E-04 & 0.13417818E-04 & 0.13415986E-04 \\
175 & 0.40993408E-06 & 0.41717685E-06 & 0.41711848E-06 \\
180 & 0.32151105E-10 & 0.16725576E-17 & 0.10545833E-16 \\
185 & 0.42435937E-06 & 0.41717540E-06 & 0.41711978E-06 
\etb
\caption{The oscillator strength around 45$^{\circ}$ and 90$^{\circ}$ for [CuCl$_4$]$^{2-}$
         in different symmetries.}
\label{cucl_tab}
\et

\subsection{Stability of K pre edge XAS intensities of [FeCl$_4$]$^{1-}$}

[FeCl$_4$]$^{1-}$ has a tetrahedral ligand environment. In T$_d$
symmetry, the metal 3d orbitals belong to the $e$ and $t_2$
irreducible representations, see Figure \ref{active_space_fig}. The
iron 4p orbitals also have $t_2$, which means that they can mix
through the interactions with the Cl ligands. The first two pre edge
transitions are to the 3d($e$) orbitals, and are electric dipole
forbidden. The next three are to the $t_2$ orbitals, and they are
more intense as they are electric-dipole allowed through the 4p mixing
and get large contributions from several orders in the multipole
expansion.\cite{westre1997multiplet,guo_2016,oiqm} Not only will the electric quadrupole $f_{0n}^{(Q^2)}$ term be large
but the electric dipole $f_{0n}^{(\mu^2)}$ will be very large
and the electric dipole electric octupole $f_{0n}^{(\mu O)}$ term
will also be significant even when the coordinate system is placed on the Fe atom.

In previous applications that examined the
origin independence of the multipole expansion in a mixed gauge certain transitions in [FeCl$_4$]$^{1-}$ gave negative oscillator strengths
at the second order despite the fact that the zeroth order in the multipole
expansion of the oscillator strengths should be the dominant term.\cite{oiqm} The
strong negative oscillator strengths, however, only appeared in the cc-pVDZ and AUG-cc-pVDZ
basis sets but not in the ANO-RCC basis sets. Since the oscillator strengths
for the exact operator is inherently positive it would therefore be interesting
to see what values the multipole expansion should converge to, and second,
to make a comparison of the numerical stability and performance of the exact operator and
the multipole expansion.


In Table \ref{sfecl4} the total dipole, quadrupole and exact intensities for the
transition from the ground to selected core-excited states in $[FeCl_4]^{1-}$ is
shown in the ANO-RCC-VTZP and AUG-cc-pVDZ basis sets. The
$G\rightarrow C1$ transition reaches the 3d($e$) orbital, while both
C3 and C5 are 3d($t_2$) final states. C12 is a two-electron excitations, with both core and valence
electrons excited simultaneously, and are typically weaker than the
main transitions.

\bt
\btb{lcccccc}
Basis & Transition & $f_{0n}^{(\mu^2)}$ & $f_{0n}^{(\mu^2)^p}$ & $f_{0n}^{(2)}$ & Total & Exact \\ \hline
ANO-RCC-VTZP & $G\rightarrow C1$  & 0.157E-12 & 0.151E-12 &  0.407E-05 &  0.407E-05 & 0.371E-05 \\
AUG-cc-pVDZ  & $G\rightarrow C1$  & 0.765E-06 & 0.447E-06 &  0.258E-05 &  0.335E-05 & 0.472E-05 \\ \hline
ANO-RCC-VTZP\cite{oiqm}& $G\rightarrow C3$ & 0.283E-04 & 0.273E-04 &  0.144E-05 &  0.427E-04 & 0.305E-04 \\
AUG-cc-pVDZ\cite{oiqm} & $G\rightarrow C3$ & 0.281E-04 & 0.168E-04 & -0.585E-04 & -0.304E-04 & 0.209E-04 \\ \hline
ANO-RCC-VTZP & $G\rightarrow C5$  & 0.283E-04 & 0.273E-04 &  0.144E-05 &  0.427E-04 & 0.305E-04 \\
AUG-cc-pVDZ  & $G\rightarrow C5$  & 0.234E-04 & 0.169E-04 & -0.539E-04 & -0.305E-04 & 0.210E-04 \\ \hline
ANO-RCC-VTZP & $G\rightarrow C12$ & 0.555E-09 & 0.419E-09 & -0.353E-09 &  0.202E-09 & 0.484E-09 \\
AUG-cc-pVDZ  & $G\rightarrow C12$ & 0.730E-08 & 0.627E-08 &  0.752E-08 &  0.148E-07 & 0.624E-08 \\
\etb
\caption{The total dipole- and quadrupole, second order and exact oscillator strengths for the
         transition from the ground (G) to selected
         core-excited (CX) state in $[FeCl_4]^{1-}$ without spin-orbit coupling. The second order (Total) is the
         sum of the electric dipole $f_{0n}^{(\mu^2)}$ 
         and the second order contribution $f_{0n}^{(2)}$ of the 
         multipole expansion. The dipole is given in both the length gauge
         $f_{0n}^{(\mu^2)}$ and velocity gauge $f_{0n}^{(\mu^2)^p}$.}
\label{sfecl4}
\et
 
For the $G\rightarrow C1$ transition, the second order contributions
($f_{0n}^{(2)}$) dominate the multipole expansion and the electric
dipole $f_{0n}^{(\mu^2)}$ approach numerical noise, see Table \ref{sfecl4}. The total
oscillator strengths in the multipole expansion are then rather similar in the two basis
sets. Instead looking at the C3 and C5 transitions, they have large
$f_{0n}^{(\mu^2)}$ contributions, which should lead to more intense
transitions than for C1. However, the presence of large electric
dipole contributions leads to large and unstable second-order
contributions, even to the point where the  total oscillator strength
becomes negative for the AUG-cc-pVDZ basis set. Finally, the C12
transition illustrates that even if the total oscillator strength is
positive, the multipole expansion leads to unphysical negative
second-order contributions.\cite{oiqm}

Instead looking at the results for the exact operator, the differences
between the basis sets are significantly smaller, even for transitions where the
second-order expansion gives total negative oscillator strengths. For transitions with strong dipole contributions the exact operator
is every time close to $f_{0n}^{(\mu^2)^p}$ which is not surprising
since we use the exact operator in the velocity gauge and the integrand
for the exact operator is closer to $f_{0n}^{(\mu^2)^p}$ than
$f_{0n}^{(\mu^2)}$.

In the ANO-RCC-VTZP we see good agreement between the second-order and exact oscillator strengths.
This is also reflected in the spectra as seen in Figure \ref{fecl4_fig}, where
only minor differences in height of the peaks can be observed. Unlike
for the multipole expansion, the peaks in the spectrum using the exact operator in the AUG-cc-pVDZ are now all 
positive. The better agreement between different basis sets could indicate that the 
exact operator is numerically more stable and reliable than the multipole expansion
though further numerical and theoretical investigation would be needed to conclude this.
Studies along these lines are currently being undertaken. 

\bef[h!]
\bc
\includegraphics[width=0.5\textwidth,angle=0]{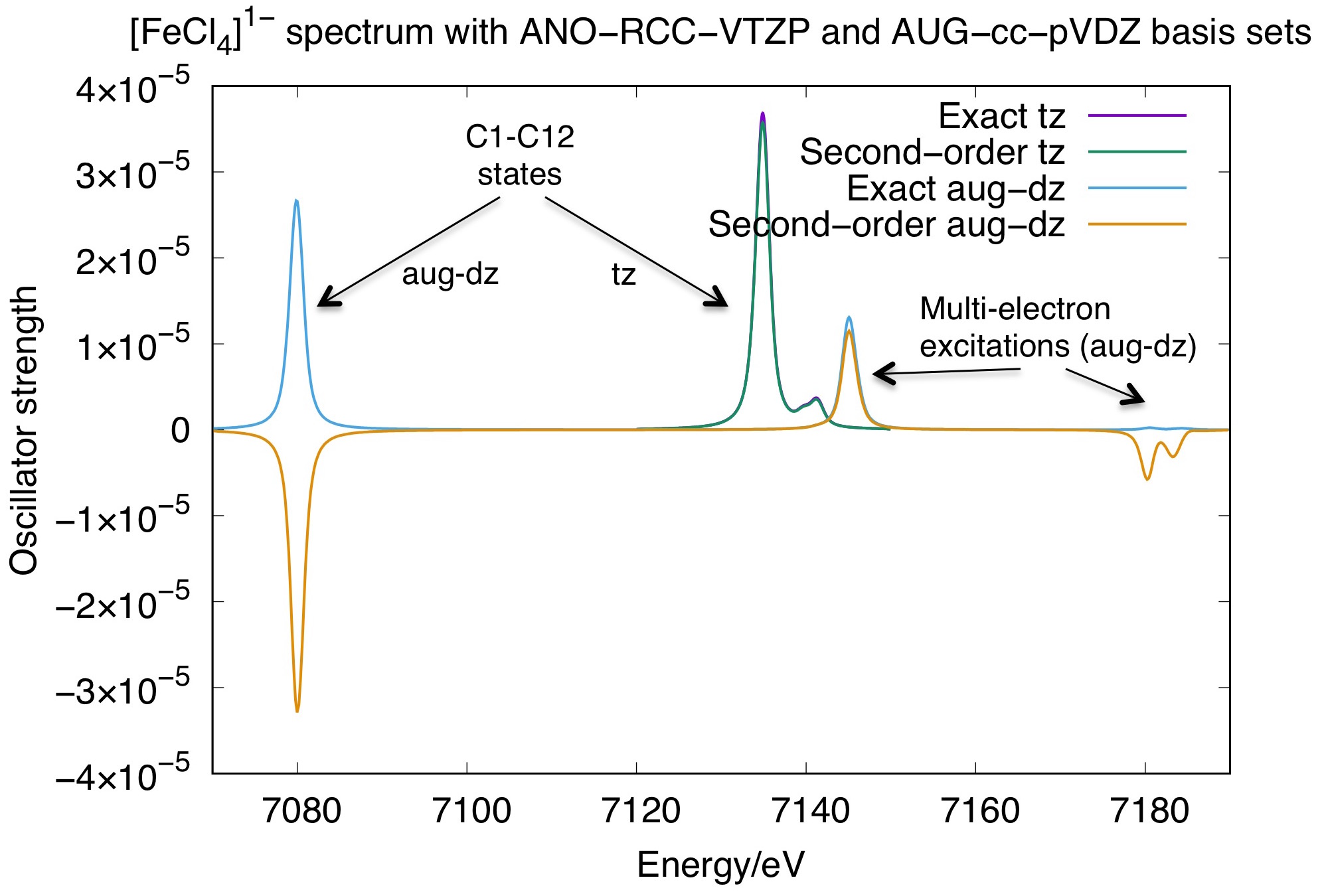}
\caption{A comparison of the spectra for $[FeCl_4]^{1-}$ 
        using the exact operator and the second-order expansion 
        in the AUG-cc-pVDZ and ANO-RCC-VTZP basis sets. Note that the
        spectra are energetically shifted due to different
        descriptions of the core orbitals. The spectra for the
        second order was previously published in Ref. \cite{oiqm}.}
\label{fecl4_fig}
\ec
\ef

\subsection{Origin dependence of metal K pre edge XAS of iron dimer}

As shown by Bernadotte \etal, the full second-order expansion is
origin independent in the velocity gauge.\cite{Lehtola,bernadotte2012origin} We showed that this also
holds in the true length gauge, but not in the mixed
gauge that is typically referred to as the length
gauge.\cite{oiqm,SORENSEN2017} This becomes an issue for an iron dimer that lacks a natural origin
for the multipole expansion. As the individual iron sites in the dimer
are asymmetric, metal 4p orbitals mix into the
valence space, giving dipole-allowed transitions
in the pre-edge, which leads to instability for the second-order
expansion.

In \cite{oiqm} we showed that if the origin was placed close to the center of mass
the change in the spectrum for the so-called origin independent quadrupole oscillator
strengths in a mixed gauge was minor while at slightly larger distances
significant changes in the spectrum could be observed, see Figure
\ref{fe2o_move_origin}. The intensity of the second peak is most
sensitive, which is consistent with larger electric-dipole
contributions.

\bef[h!]
\bc
\includegraphics[width=0.5\textwidth,angle=270]{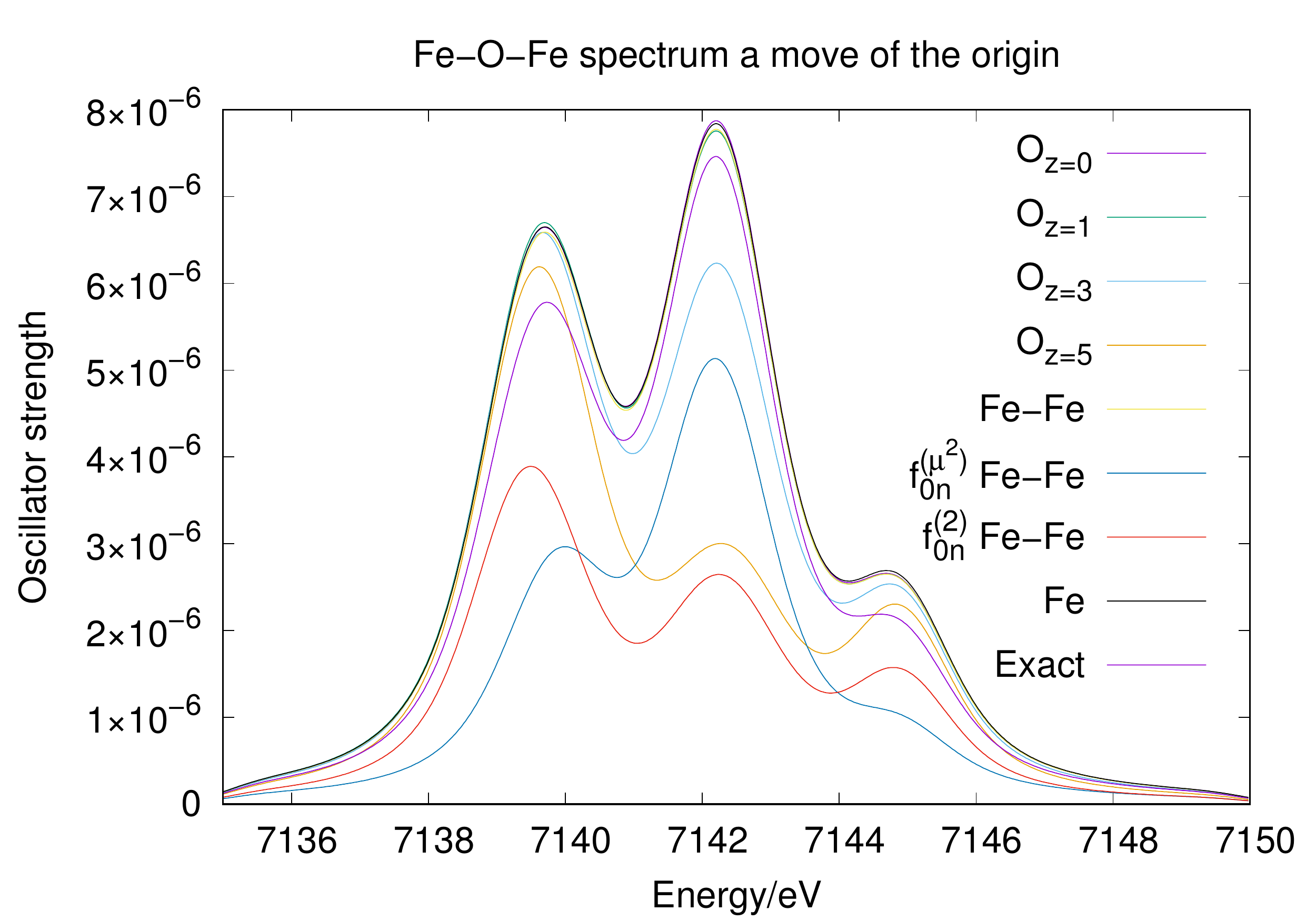}
\caption{A comparison of the spectra for Fe-O-Fe with the origin in the
         oxygen atom (O$_{x=0}$), origin moved along the $x$-axis (O$_{x=d}$),
         where $d$ is the distance,
         the origin in the middle between the two Fe atoms (Fe-Fe)
         and the origin placed on one of the Fe atom (Fe) and the exact operator
         in the ANO-RCC-VTZP basis set. The $f_{0n}^{(\mu^2)}$ and
         $f_{0n}^{(2)}$ contributions are shown with the origin
         placed between the two Fe atoms. The data from the 
         origin independent quadrupole oscillator strengths in a mixed gauge
         is taken from Ref. \cite{oiqm}.}
\label{fe2o_move_origin}
\ec
\ef

From Figure \ref{fe2o_move_origin} we see that the exact operator and the 
oscillator strengths in the mixed gauge
agree rather well, both with respect to the shape and the total intensity of the 
spectrum. Previous K pre-edge calculations using the mixed gauge are
therefore most likely of acceptable quality.

%% file: per.tex
While the results in Sec. \ref{SEC:appl} and implementation in 
Sec. \ref{eva} does show that the exact semi-classical
light-matter interaction is easier to implement and
numerically better than the multipole expansion for the
weak field limit we believe that the real strength of
the approach lies in the strong field regime.

\subsection{Real-time dependent light matter interaction}
\label{real_time}

For strong fields, where the perturbative treatment of the 
light-matter interaction breaks down, the multipole expansion
is still used and the light-matter interaction is usually treated
in the dipole approximation 
\beq
\hat H = \hat H_0 -  \bm{E}(t) \bm{\mu}.
\eeq
If the wave function is expanded in a Gaussian basis then
the exact same as evaluation of the exact semi-classical light
matter interaction presented in Section \ref{eva} could be used 
without any significantly added cost to a more general
Hamiltonian
\beq
\hat H = \hat H_0 + \hat U(t),
\eeq
where $\hat U(t)$ is given by Eq. \ref{tinde},
to significantly improve
the description of a laser-pulse interacting with a target.
Going beyond the dipole approximation is particularly 
interesting for X-ray spectroscopy and in general 
for very short wave lengths where the
pulse varies over the size of the molecule, very strong
time-dependent magnetic fields, for multi-photon processes
where the field becomes strong enough to see contributions 
from the $(\bm{A}^2)$ in Eq. \ref{tinde2}. While the differences in
oscillator strengths seems minor with well-behaving basis sets in the
static case, see Figure \ref{fe2o_move_origin}, these differences should quickly become apparent in the dynamic case
where it is the interaction with the laser field that drives the dynamics
since small initial differences in interaction can quickly grow large.
In the real-time dependent description the explicit strength and shape
of the field would also have to be included though 
these are merely the values of a time dependent function
describing the envelope and strength of the field.

%% file: summary.tex
We have presented a very easy way to implement the exact
semi-classical light-matter interaction from Eq. \ref{tinde}
where the integrals either can be calculated analytically
or extending the standard Gauss-Hermite integral evaluation
to complex numbers. 
We show that the integral evaluation is akin to a Fourier transformation
from real to $k$-space of the overlap of the basis functions and that the electric, magnetic
and quadratic term $(\bm{A}^2)$ can be evaluated in
the same way.

The main advantages of the exact operator is seven-fold: {\it i}) it
is cheaper to calculate than higher order terms
in the multipole expansion, {\it ii}) there is never negative
oscillator strengths, {\it iii}) always origin independent, 
{\it iv}) easy to extend to time-dependent calculations, {\it v})
appears to be more numerically stable, and {\it vi}) less sensitive 
to the choice of basis set since the
basis set only have to work for a single type of integrand
and not a multitude of different integrands as in the multipole expansion \cite{crossley}.
Additionally, {\it vii}) using the exact operator also avoids the faulty
interpretation of electric and magnetic terms in the
multipole expansion since this interpretation always
will depend on the choice of coordinate system since
none of these terms are observable.
Due to the ease at which the exact semi-classical light matter interaction can be
implemented, the numerical, theoretical and interpretation 
advantages we do not see the need for the multipole expansion
anymore for transition moments.

We show numerical examples of the exact operator on [CuCl$_4$]$^{2-}$ where
the angle between beam and sample is known
and on [FeCl$_4$]$^{1-}$ and Fe-O-Fe where an isotropic averaging is performed.
For the bis(creatinium)tetrachlorocuprate(II) crystal 
we have showed that with an angular resolved spectrum can
discern the symmetry of the [CuCl$_4$]$^{2-}$ unit.
The numerical stability of the exact operator,
even in basis sets performing poorly for the multipole expansion,
has been demonstrated for the [FeCl$_4$]$^{1-}$ molecule.
In the AUG-cc-pVDZ basis set even transitions with negative second order
oscillator strengths the exact operator gave results
close to those obtained for the exact operator in the better ANO-RCC-VTZP basis set where
the second order oscillator strengths was positive.
In fact the difference between the oscillator strengths for
the exact operator in the ANO-RCC-VTZP and AUG-cc-pVDZ basis sets
is about the same as the difference between the 
exact operator and the second order in the ANO-RCC-VTZP basis set.
If the very good numerical performance of the exact operator,
seen in these preliminary calculations, is general is currently being
investigated.
Finally for Fe-O-Fe we reproduce the spectrum previously published\cite{oiqm}
which together with the results for [FeCl$_4$]$^{1-}$ in the ANO-RCC-VTZP basis
shows that when good basis sets are used then the
multipole expansion does produce results close
to that of the exact operator.

While using the exact operator does give a significant 
improvement over the multipole expansion for weak fields
we do believe that the real strength of the approach
will be in the dynamics of strong fields.
We are therefore currently exploring the options of using the
exact operator in time-dependent calculations since
this will give more accurate dynamics for molecules in strong
laser fields, here in particularly for very short wave lengths,
like X-rays, where the field varies over the range of
the molecule or an atom, where terms above the
dipole becomes important or where the $\boldmath{A}^2$ becomes important.

The implementation of the exact operator for electric and magnetic
fields along with the integrals the for quadratic term $(\bm{A}^2)$ 
are freely available in OpenMolcas.\cite{OpenMolcas}

%% file: ack.tex
Financial support was received
from the Knut and Alice Wallenberg Foundation for the
project “Strong Field Physics and New States of Matter”
(Grant No. KAW-2013.0020) and the Swedish Research Council (Grant No. 2012-3910, 2012-3924, and 2016-03398).
Computer resources were provided by SNIC trough the National Supercomputer Centre at Link{\"o}ping University (Triolith) under projects snic2014-5-36,  snic2015-4-71, snic2015-1-465 and snic2015-1-427.